\documentclass[twocolumn]{aastex631}

\usepackage{amsmath}	
\graphicspath{{figures/}}
\newcommand{\pcm}{\textrm{pc\,cm}$^{-3}$}

\usepackage[normalem]{ulem}

\shorttitle{Propagation effects and FRBs}
\shortauthors{Kumar et al.}

\begin{document}

\title{Impact of propagation effects on the spectro-temporal properties of Fast Radio Bursts}

\author[0009-0006-2182-1440]{Aishwarya Kumar}
\affiliation{Department of Physics and Astronomy, The University of Western Ontario, 1151 Richmond Street,London, Ontario N6A 3K7, Canada}

\author[0000-0002-1453-7074]{Fereshteh Rajabi}
\affiliation{Department of Physics and Astronomy, McMaster University, 1280 Main Street West, Hamilton, Ontario, L8S 4L8, Canada}

\author[0000-0003-4420-8674]{Martin Houde}
\affiliation{Department of Physics and Astronomy, The University of Western Ontario, 1151 Richmond Street,London, Ontario N6A 3K7, Canada}

\begin{abstract}

We present a mathematical analysis of propagation-induced distortions in the spectro-temporal properties of Fast Radio Bursts (FRBs). Within the Triggered Relativistic Dynamical Model, we derive a centroid-based formulation of the sub-burst slope law, which is an inverse relation between the frequency drift rate and the temporal width of sub-bursts. We extend this framework to include two frequency-dependent propagation effects: \textup{(i)} multipath scattering, characterized by a pulse-broadening timescale $\tau_\mathrm{sc} \propto \nu^{-4}$, and \textup{(ii)} residual dispersion, parameterized by $\Delta \mathrm{DM}\propto \nu^{-2}$. Our analysis shows that scattering preserves the inverse relation between sub-burst slope and duration, but increases the scaling coefficient when $\tau_\mathrm{sc}$ exceeds the intrinsic width ($t_\mathrm{w}$) of sub-bursts. Residual DM errors act asymmetrically: under-dedispersion flattens the sub-burst slope, whereas over-dedispersion causes a non-linear increase and eventually a change of sign. When both effects are present, scattering counterbalances the steepening induced by over-dedispersion and augments the flattening produced by under-dedispersion, yielding characteristically distorted curves. We repeat measurements for ultra-short duration bursts (ultra-FRBs) with $t_\mathrm{w} = 50\  \mu\mathrm{s}$ at 1 GHz and find them to be far more sensitive to propagation errors. Deviations become measurable for $\left | \Delta \mathrm{DM} \right |\sim0.05$~\pcm~and for $\tau_\mathrm{sc} \sim0.1$ ms at 1 GHz, levels that have negligible impact on the standard-width sub-bursts. Our analysis provide practical diagnostics to disentangle propagation effects from the observed spectro-temporal properties of FRBs, thereby recovering true correlations among their intrinsic parameters.

\end{abstract}


\section{Introduction} \label{sec:intro}

Fast Radio Bursts (FRBs) are extraordinarily bright, short-duration transients, generally of extragalactic origin \citep{Lorimer_2007, Review_ppr_2019, cordes2019fast}. They are typically categorized on the basis of their activity rates into two distinct subtypes: repeating and non-repeating bursts. FRBs exhibit diverse spectro-temporal characteristics, energy distributions, periodicity (for repeating FRBs), and polarization properties \citep{Zhang_2024}. Despite extensive observations, the origins and emission mechanisms of FRBs remain unclear. Establishing correlations between their complex and varied properties serves as a powerful instrument to construe the causation of these events. One such observable used for probing the underlying emission mechanism of FRBs is the sub-burst slope\footnote{Following established terminology, we refer to a ``sub-burst'' as a temporally and spectrally localized component within the dynamic spectrum of an FRB. The ``sub-burst slope law'' applies to individual sub-bursts, while the ``drift law'' refers to bursts containing multiple sub-bursts \citep{Chamma2021,Chamma2023}.}. It quantifies the frequency-dependent arrival time delay within a single sub-component of an FRB event. The sub-burst slope law describes the functional dependence of this slope on either the observing frequency or the sub-burst duration, and is a characteristic feature of the Triggered Relativistic Dynamical Model (TRDM) introduced by \citet{Rajabi2020}.

As bursts propagate from the source to the observer, their spectro-temporal profiles undergo modifications due to propagation effects such as dispersion, scattering, and scintillation. Extracting dispersive delays and scattering timescales is both challenging and crucial for the analysis of the intrinsic properties of FRBs. The underlying dispersion measure (DM) offers insights into the aggregate electron number density encountered along the path. Different methods allow us to estimate the DM and remove it using de-dispersion techniques \citep{Review_ppr_2019}. However, inaccuracies in these estimates can lead to residual dispersion in a burst, thereby affecting the measurement of its properties. Scattering, caused by irregularities in the electron density \citep{Scheuer1968, Rickett1977, Cordes2002}, adds temporal smearing, which is often identified by an exponential tail observed in the frequency-integrated burst profile. Numerous studies have identified and quantified scattering in both repeating and non-repeating FRBs \citep{Shannon2018, Ravi2019, Farah2019, CHIME2019_Amiri, Ocker_2022, Ocker2023,  CHIME2023_Andersen}. While sub-burst slope analyses are limited to repeaters, these works provide empirical constraints on the scattering timescales explored in our study.

This paper investigates the effects of scattering and inaccurate de-dispersion on the spectro-temporal features of FRBs by analyzing deviations in the sub-burst slope law. We begin with an overview of relevant propagation effects, the TRDM, and the sub-burst slope law in Section \ref{sec:theory}. Section \ref{sec:math} develops a centroid-based formulation of the slope law and incorporates scattering and residual-DM terms, treating the two effects both separately and in conjunction. Section \ref{sec:res} presents the resulting spectro-temporal modifications for a range of scattering timescales and DM offsets for both standard-width and ultra-FRBs. In Section \ref{sec:dis}, we quantify and interpret the shifts produced by these propagation effects and discuss their observational implications. A summary of our findings is provided in Section \ref{sec:sum}.

\section{Propagation effects and the Sub-Burst Slope Law}
\label{sec:theory}

\subsection{Propagation effects }
\label{sec:prop_eff}
The non-homogeneous distribution of electron density in galaxies leads to sub-bursts having multiple propagation paths, resulting in differential arrival times for signal components and temporal smearing of the pulse shape. While the microscopic scattering process is governed by stochastic fluctuations in plasma density, its effect on the pulse profile can be described statistically by a characteristic scattering timescale, $\tau_{\mathrm{sc}}$, which depends on the observing frequency, $\nu$, as follows:
\begin{align}
   \tau_\mathrm{sc} = \Lambda_\mathrm{sc}\left(\frac{\nu}{1\,\mathrm{GHz}}\right)^{-n}, 
   \label{eq:ts}
\end{align}
where $n=4.0$ for the thin screen model and $n=4.4$ for the Kolmogorov spectrum \citep{Rickett1977}. The constant of proportionality, $\Lambda_\mathrm{sc}$, depends on the scale size of the irregularities, the magnitude of the electron density fluctuations, and the distance of the source from the observer. Although we adopt $\Lambda_\mathrm{sc}$ values between 0 to 20 ms for our analysis, bursts with scattering timescales outside this range have also been observed \citep{Ravi2019}. 

As the pulse travels through different environments, the ionized components around the source, the intergalactic medium, and the Milky Way introduce dispersion in its spectra. Dispersion is a frequency-dependent delay that causes the lower-frequency components of a sub-burst to arrive later than the higher-frequency components. This time delay at frequency $\nu$ is expressed as
\begin{align}
   \Delta t = a \, \mathrm{DM} \left ( \frac{1}{\nu^2} - \frac{1}{\nu_{\mathrm{ref}}^2} \right ),
   \label{eq:t_dm}
\end{align}
where $a=4.148\,806\,4239(11)\;\textrm{GHz}^2\,\textrm{cm}^3\,\textrm{pc}^{-1}\,\textrm{ms}$ \citep{Kulkarni2020} and $\nu_{\textrm{ref}}$ is a reference frequency, typically set to the highest frequency present in a dynamic spectrum or to infinity. 

Low signal-to-noise ratio (S/N) and insufficient time resolution make it harder to decouple the intrinsic sub-burst spectra from the propagation effects, leading to imprecise measurements of scattering timescales and DM. As $\tau_\mathrm{sc}$ depends on stochastic electron density fluctuations along the line-of-sight, it can vary between bursts for the same repeater \citep{Ocker_2022}. Statistical DM uncertainties are often quoted at the $\lesssim 1\%$-level of the reported DM value. For instance, FRB 20191221A has a DM of 368~\pcm~with an uncertainty of $\sigma_\mathrm{DM} \simeq 6$~\pcm~\citep{CHIME2022_Andersen}. Yet, the published DM values for a single source can differ because they depend upon the timing of observation, the instrumentation used, the specific de-dispersion pipeline employed, and the metric optimized to select the DM (e.g., based on the S/N or the structure of the burst). \citet{Chamma2021, Chamma2023} and \citet{Brown2024} determine the DM for FRB sources by identifying the value that best fits the sub-burst slope law. While most representative DMs calculated through this approach are typically consistent with cataloged values, there are some outliers. For example, the reported DM for FRB 20180301A in \citet{Price2019} is $522\pm5$~\pcm. \citet{Brown2024} found that some sub-bursts were over-corrected at this value, resulting in non-physical positive slopes, according to the TRDM. They found the representative DM to be 515.4~\pcm, resulting in a discrepancy of $\sim7$~\pcm. Such incongruities in estimating DM and scattering timescales introduce frequency-dependent distortions in the dynamic spectrum across the observing band and can bias the measurements of spectro-temporal properties of FRBs.

\subsection{The Sub-burst Slope Law }
\label{sec:trdm}

Within the framework of the TRDM, an FRB source is modeled as consisting of multiple components that move, potentially at relativistic speeds, relative to the observer \citep{Houde_2018, Houde2019,Rajabi2020}. Following a triggering event originating from a background source, each component of the FRB source emits narrow-band radiation after a time delay. Due to the relativistic Doppler shift, radiation emitted at a frequency $\nu^\prime$ in the source's rest frame is detected at frequency $\nu$ in the observer's frame. The finite velocity distribution covered by the components of the FRB source, coupled with the relativistic Doppler shift, transforms the individual narrow-band spectra into the wide-band emission detected as a sub-burst. \citet{Rajabi2020} expressed the sub-burst slope law as

\begin{align}
        \frac{1}{\nu}\frac{d \nu}{dt_\mathrm{D}} =-\left (  \frac{\tau_\mathrm{w}^\prime}{\tau_\mathrm{D}^\prime}\right )\frac{1}{t_\mathrm{w}}=-\frac{A}{t_\mathrm{w}},
   \label{eq:slplaw}
\end{align}

\noindent where $A$ denotes the sub-burst slope parameter, which encapsulates intrinsic source properties and is expressed as a function of the proper delay ($\tau_\mathrm{D}^\prime$) and proper duration ($\tau_\mathrm{w}^\prime$). These rest-frame quantities are related to their observer-frame counterparts, $t_\mathrm{D}$ and $t_\mathrm{w}$, through the following transformations: 

\begin{align}    
   t_\mathrm{D} & =\tau_\mathrm{D}'\sqrt{\frac{1-\beta}{1+\beta}}=\tau_\mathrm{D}'\frac{\nu^\prime}{\nu}
   \label{eq:td},\\
   t_\mathrm{w} & =\tau_\mathrm{w}'\sqrt{\frac{1-\beta}{1+\beta}}=\tau_\mathrm{w}'\frac{\nu^\prime}{\nu}.
   \label{eq:tw}
\end{align}

\noindent Here, $\beta$ is the velocity (divided by the speed of light) of the FRB source relative to the observer. The analyses of \citet{Rajabi2020}, \citet{Chamma2021}, \citet{Jahns2023}, \citet{Chamma2023}, and \citet{Brown2024} provide observational evidence for the aforementioned relationships. By leveraging the sub-burst slope law, \citet{Chamma2023} and \citet{Brown2024} derive representative DMs for the sub-bursts. These studies also found that the value of $A$ for individual sources can differ from that obtained when fitting multiple sources simultaneously. While the method remains robust in general, its validity must be evaluated in the presence of propagation effects, and the consequent perturbations to $A$ should be assessed quantitatively.

It is critical to acknowledge that measurements of the sub-burst slope differ depending on the analytical method employed \citep{Gopinath_2023}. Methods based on two-dimensional elliptical Gaussian fitting, applied either directly to the burst \citep{Jahns2023} or to its auto-correlation function \citep{Chamma2021,Chamma2023,Brown2024}, preserve the scattering imprint and are sensitive to residual dispersion. As a result, the measured sub-burst slopes depend explicitly on both the scattering timescale and the assumed dispersion measure, rather than on the intrinsic emission alone. In contrast, methods based on time-of-arrival (TOA) estimates \citep{Gopinath_2023, Chamma_2025} avoid incorporating the entire profile but are prone to TOA errors, such as those arising from inaccurate de-dispersion.

\section{Analytical formulation of the sub-burst slope law with scattering and residual dispersion}
\label{sec:math} 

\subsection{General Methodology}
\label{sec:gen_methods} 

We model the sub-burst intensity profile as a decaying exponential:
\begin{align}
    I\left(\nu, t\right)=\frac{F_0}{t_\mathrm{w}}\exp \left [ {-\frac{\left(t-t_\mathrm{D}\right)}{t_\mathrm{w}}} \right ]H\left(t-t_\mathrm{D}\right),
	\label{eq:in_1}
\end{align}
where $t_\mathrm{w}$ and $t_\mathrm{D}$ are the intrinsic duration and delay, respectively (as defined in Equations \ref{eq:tw} and \ref{eq:td}), $F_0$ is the fluence, and $H\left(t - t_\mathrm{D}\right)$ is the Heaviside distribution. From now on, we will drop the Heaviside function in future definitions of the sub-burst profile, with the implicit understanding that the signal is zero for $t<t_\mathrm{D}$. Equation (\ref{eq:in_1}) represents an unaltered sub-burst, devoid of any propagation effects. Its exact functional form will vary depending on the propagation effect under consideration and is discussed in the subsequent sections.  We adopt an exponentially decaying function, as it ensures both convergence and analytical tractability of the integrals involved in determining the parameters used in this analysis. Nonetheless, the methodology outlined below remains applicable to other sub-burst profiles.

After establishing a profile, the temporal centroid, $t_\mathrm{c}$, of a sub-burst at frequency $\nu$ is defined by evaluating its first moment, 
\begin{align}
     t_\mathrm{c}  = \frac{\int_0^\infty t \cdot I(\nu ,t) \ dt}{\int_0^\infty  I(\nu ,t) dt}.
    \label{eq:t_c_gen}
\end{align}
As $\nu$ varies across the sub-burst bandwidth, these centroid times trace out a frequency-dependent trajectory, given by $t_\mathrm{c}(\nu)$. The sub-burst duration (or width) is obtained from the second central moment of the intensity profile about this centroid:
\begin{align}
     \lambda (\nu) =  \frac{\int_0^\infty (t - t_\mathrm{c})^2 \cdot I(\nu ,t) \ dt}{\int_0^\infty  I(\nu ,t) dt}.
    \label{eq:dur_gen}
\end{align}
Evaluating this equation at the central frequency yields $ \lambda(\nu_0)\equiv \lambda_\mathrm{c}$, which we adopt as the representative duration of the sub-burst.

Subsequently, we compute the frequency derivative of the centroid time trajectory, which is crucial for measuring the sub-burst slope:
\begin{align}
     \frac{dt_\mathrm{c}}{d \nu} = \frac{d}{d \nu} \left [ \frac{\int_0^\infty t \cdot I(\nu ,t) \ dt}{\int_0^\infty  I(\nu ,t) dt} \right ].
   \label{eq:dtc_dnu_gen}
\end{align}
Finally, following \citet{Rajabi2020}, the frequency-normalized sub-burst slope measured relative to the centroid is expressed as:
\begin{align}
    \left \langle \frac{1}{\nu} \frac{d \nu}{dt_\mathrm{c}} \right \rangle_{\nu_0} = \frac{1}{\Delta \nu}\int_{\nu_0 - \Delta \nu/2}^{\nu_0 + \Delta \nu/2} \left ( \frac{1}{\nu} \frac{d\nu}{dt_\mathrm{c}} \right ){d\nu} .
   \label{eq:slp_law_tc_gen}
\end{align}
where $\Delta \nu$ is the bandwidth of the sub-burst centered around frequency $\nu_0$.

\begin{figure}\centering
 \includegraphics[scale=0.5]{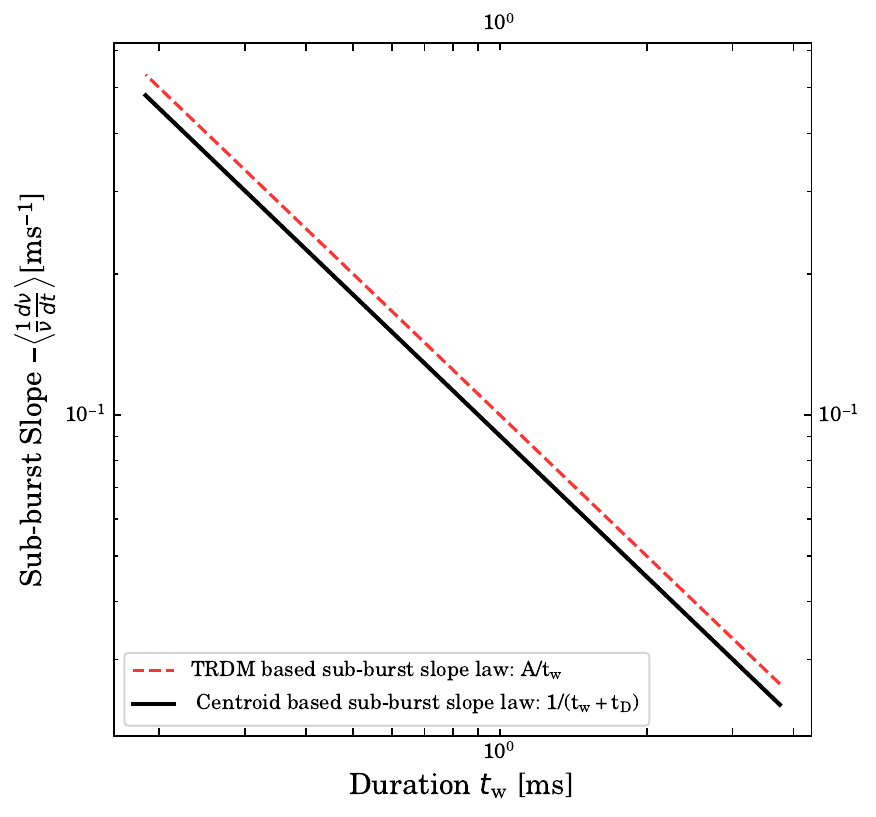}
 \caption{Comparison of two formulations of the sub-burst slope laws: the dashed red line represents the sub-burst slope law when measured relative to the time of arrival $t_\mathrm{D}$ (Equation \ref{eq:slplaw}), whereas the solid black line corresponds to the sub-burst slope law measured relative to the centroid time $t_\mathrm{c}$ (Equation \ref{eq:slp_law_tc}).}
 \label{fig:slp_law_comp}
\end{figure}

Applying this procedure to the intensity profile in Equation (\ref{eq:in_1}) yields the sub-burst slope law
\begin{align}
    \left \langle \frac{1}{\nu} \frac{d \nu}{dt_\mathrm{c}} \right \rangle_{\nu_0} = -\frac{1}{\Delta \nu}\int_{\nu_0 - \Delta \nu/2}^{\nu_0 + \Delta \nu/2}\frac{1}{t_\mathrm{D} + t_\mathrm{w}} {d\nu},
   \label{eq:slp_law_tc}
\end{align}
with $t_\mathrm{c} = t_\mathrm{D} + t_\mathrm{w}$. This ideal centroid-based sub-burst slope law lies below the ideal TRDM law (Equation \ref{eq:slplaw}) when plotted against the duration $\lambda_\mathrm{c} = t_\mathrm{w}$ (derived using Equation \ref{eq:t_c_gen}), as shown in Figure \ref{fig:slp_law_comp}. This is because when measuring the sub-burst drift relative to the TOA, we evaluate the change in frequency, $\Delta \nu$, with respect to the change in intrinsic delay, $\Delta t_\mathrm{D}$, of the sub-burst. Measurement of the sub-burst drift relative to the centroid records the same $\Delta \nu$ but now over a larger temporal interval, $\Delta t_\mathrm{c}$. As $\left|\Delta t_\mathrm{c}\right| > \left|\Delta t_\mathrm{D}\right|$, the sub-burst slope will be shallower when evaluated at the centroid. The overall law, however, retains its linear form in the log–log plot, as both $\Delta t_\mathrm{c}$ and $\Delta t_\mathrm{D}$ share the same $\nu^{-1}$ dependence.

\subsection{Scattering-Exclusive Formalism}
\label{sec:scat_ex}
To investigate purely scattering conditions, we model the scattering kernel as a one-sided exponential function under the thin-screen approximation \citep*{Cronyn1970, Rickett1977, Jankowski2023},
\begin{align}
    S\left(\nu, t\right)=\frac{1}{\tau_\mathrm{sc}}\exp \left ( {-\frac{t}{\tau_\mathrm{sc}}} \right ) H\left(t\right),
	\label{eq:sf}
\end{align}
where $\tau_\mathrm{sc}$ denotes the scattering timescale. The post-scattering sub-burst profile, resulting from the convolution of the intrinsic intensity profile (Equation \ref{eq:in_1}) with the scattering kernel (Equation \ref{eq:sf}), is given by
\begin{align}
    I_s\left(\nu, t\right) &= \frac{F_0}{\tau_\mathrm{sc} - t_\mathrm{w}} \left\{ \exp \left[ \frac{-(t - t_\mathrm{D})}{\tau_\mathrm{sc}} \right] - \exp \left[ \frac{-(t - t_\mathrm{D})}{t_\mathrm{w}} \right] \right\}.
	\label{eq:con_fin}
\end{align}

From Equation (\ref{eq:t_c_gen}), the first temporal moment of this intensity profile yields the centroid:
\begin{align}
     t_\mathrm{c}
    = t_{\mathrm{D}} + t_{\mathrm{w}}+  \tau_\mathrm{sc}
    \label{eq:tc_scat}
\end{align}

Empirical studies report $t_\mathrm {w}\propto\nu^{-1}$ \citep{Chamma2023,Brown2024}, while the TRDM relates $t_\mathrm {D}$ and $t_\mathrm {w}$ through $t_\mathrm {D}=t_\mathrm {w}/A$, with $A\approx 0.1$.  The frequency dependence of $\tau_\mathrm{sc}$ is expressed in Equation (\ref{eq:ts}). Leveraging these relations, we evaluate the derivative of $t_\mathrm{c}$ with respect to frequency
\begin{align}
     \frac{dt_\mathrm{c}}{d\nu} = -\frac{1}{\nu} \left [ t_\mathrm{D} + t_\mathrm{w} + n\tau_\mathrm{sc}\right ],
	\label{eq:tc_dnu}
\end{align}
where $n$ is the scattering index. Averaging Equation (\ref{eq:tc_dnu}) across the sub-burst bandwidth $\Delta \nu$, centered on $\nu_0$, yields the frequency-normalized sub-burst slope:
\begin{align}
     \left \langle \frac{1}{\nu}\frac{d\nu}{dt_{\mathrm c}}\right \rangle_{\nu_0}
   = -\frac{1}{\Delta\nu}
     \int_{\nu_0 - \Delta\nu/2}^{\nu_0 + \Delta\nu/2}
         \frac{\,{d\nu}}{t_{\mathrm D} + t_{\mathrm w} + n\tau_\mathrm{sc}}.
	\label{eq:slp_law_scat}
\end{align}

We analytically ascertain the characteristic duration of the scattering-exclusive profile through Equation (\ref{eq:dur_gen}):
\begin{align}
     \lambda_\mathrm{c} =  \sqrt{t_{\mathrm{w}}^2 + \tau_\mathrm{sc} ^2}.
    \label{eq:lam_scat}
\end{align}

The duration $\lambda_\mathrm{c}$ inherits the explicit frequency dependence of both the intrinsic sub-burst duration $ {t_\mathrm{w}(\nu_0)}$ and the scattering timescale $ {\tau_\mathrm{sc}(\nu_0)}$ (Equations \ref{eq:tw} and \ref{eq:ts}, respectively), thereby encapsulating the corresponding influence of scattering.

\subsection{DM-Exclusive Formalism}
\label{sec:dm_ex}

We begin by defining the frequency-dependent residual dispersive delay in a sub-burst at frequency $\nu$ as
\begin{align}
   \Delta t_{\mathrm{DM}}= a \,\Delta\mathrm{DM} \left (\frac{1}{\nu^2} -\frac{1}{\nu^2_\mathrm{ref}} \right ) \: \: \mathrm{ms},
	\label{eq:dm_1}
\end{align}
following Equation (\ref{eq:t_dm}). In the subsequent analysis, $\Delta\textrm{DM}$ refers to the residual DM remaining after de-dispersion,
\begin{align}
    \Delta\mathrm{DM}= \mathrm{DM}_\mathrm{true}-\mathrm{DM}_\mathrm{est}.
    \label{eq:del_dm}
\end{align}
Here, $\mathrm{DM}_\mathrm{true}$ denotes the intrinsic DM of the source, while $\mathrm{DM}_\mathrm{est}$ is the DM estimated using a specific method or analysis. Thus, the value of $\Delta\textrm{DM} = 0$~\pcm~is indicative of perfect de-dispersion or the absence of residual dispersion in a sub-burst. A positive $\Delta \mathrm{DM}$ ($>0\,$~\pcm) implies under-dedispersion and a negative $\Delta \mathrm{DM}$ ($<0\,$~\pcm) corresponds to the case of over-dedispersion. 

To evaluate the delay in each channel, we are free to set 
$\nu_\mathrm{ref}  \rightarrow \infty $ in Equation (\ref{eq:dm_1}). We thus define our intensity function as 
\begin{align}
    I_\mathrm{DM}(\nu, t)= \frac{F_0}{t_\mathrm{w}}\exp \left [\frac{-(t-t^{\ast }_\mathrm{D})}{t_\mathrm{w}} \right ] ,
	\label{eq:in_dm}
\end{align}
where the new delay, $t^\ast _\mathrm{D}$, is due to the intrinsic delay introduced in the model ($t_\mathrm{D}$) and the delay due to dispersion ($\Delta t_{\mathrm{DM}}$), i.e., $t^\ast _\mathrm{D} =t_\mathrm{D}+\Delta t_{\mathrm{DM}}$.

Using Equation (\ref{eq:t_c_gen}), we evaluate the temporal centroid of this intensity profile:
\begin{align}
     t_\mathrm{c}  =
     t_{\mathrm{D}}+t_{\mathrm{w}} + \Delta t_\mathrm{DM}.
    \label{eq:tc_dm}
\end{align}
By substituting this in Equation (\ref{eq:slp_law_tc_gen}), we obtain the frequency-normalized sub-burst slope law as follows:
\begin{align}
     \left \langle \frac{1}{\nu}\frac{d\nu}{dt_{\mathrm c}}\right\rangle_{ \nu_0}
   = -\frac{1}{\Delta\nu}
     \int_{ \nu_0 - \Delta\nu/2}^{ \nu_0 + \Delta\nu/2}
         \frac{\,{d\nu}}{ t_{\mathrm D} + t_{\mathrm w} + 2\Delta t_\mathrm{DM}}.
	\label{eq:slp_law_dm}
\end{align}

Applying Equation (\ref{eq:dur_gen}) to the dispersion-only profile $I_{\mathrm{DM}}(\nu, t)$ provides the sub-burst duration:
\begin{align}
     \lambda_\mathrm{ c} =  t_{\mathrm{w}}({\nu_0}).
    \label{eq:lam_dm}
\end{align}
Thus, residual dispersion merely translates the burst in time without changing its intrinsic width.

\subsection{Joint Dispersion–Scattering Formalism}
\label{sec:all_in}

In the presence of both multipath scattering and residual dispersion, the resulting intensity profile takes a more complex functional form:
\begin{align}
    I_{\mathrm{joint}}(\nu , t) &= \frac{F_0}{\tau_\mathrm{sc} - t_\mathrm{w}} \left\{ \exp \left[ \frac{-(t - t^\ast_\mathrm{D})}{\tau_\mathrm{sc}} \right] \right. \nonumber\\
    &\quad - \left. \exp \left[ \frac{-(t - t^\ast_\mathrm{D})}{t_\mathrm{w}} \right] \right\}.
    \label{eq:i_joint}
\end{align}
Just as in the DM-exclusive case, we have $t^\ast _\mathrm{D} =t_\mathrm{D}+\Delta t_{\mathrm{DM}}$. Direct evaluation of the first moment yields:
\begin{align}
     t_\mathrm{c}  = t_{\mathrm{D}}+ t_{\mathrm{w}} + \Delta t_\mathrm{DM} + \tau_\mathrm{sc}.
    \label{eq:tc_joint}
\end{align}
Differentiating this equation and substituting the result into the sub-burst slope formalism, Equation (\ref{eq:slp_law_tc_gen}), gives
\begin{align}
     \left\langle \frac{1}{\nu}\frac{d\nu}{dt_{\mathrm c}}\right\rangle_{ \nu_0}
   = -\frac{1}{\Delta\nu}
     \int_{\nu_0 - \Delta\nu/2}^{\nu_0 + \Delta\nu/2}
         \frac{\, {d\nu}}{ t_{\mathrm D} + t_{\mathrm w} + 2\Delta t_\mathrm{DM} + n\tau_\mathrm{sc}}.
	\label{eq:slp_law_joint}
\end{align}

The duration of the joint profile $I_{\mathrm{joint}}(\nu , t)$ integrates exactly to be
\begin{align}
     \lambda_\mathrm{ c} =  \sqrt{ t_{\mathrm{w}}(\nu_0)^2 + \tau_{\mathrm{sc}}(\nu_0)^2},
    \label{eq:lam_joint}
\end{align}
confirming that scattering alone broadens the burst beyond its intrinsic width $t_\mathrm{w}$, while residual dispersion leaves the duration unchanged.

\section{Results}
\label{sec:res}
In the observational literature, the sub-burst slope law has been tested extensively for bursts emanating from FRB 20121102A. Studies by \citet{Rajabi2020}, \citet{Jahns2023}, \citet{Chamma2023}, and \citet{Chamma_2025} demonstrate an inverse relationship between the sub-burst slope and the duration, of the form $At_\mathrm{w}^{-1}$, as presented in Equation (\ref{eq:slplaw}). The constant $A$ was found to lie between $0.07$ and $0.1$ with the assumption that the bursts have minimal amounts of residual scattering and/or dispersion. This finding is corroborated by \citet{Chamma2021} and \citet{Brown2024}, where a similar range for $A$ is reported across multiple sources. This consistency suggests that the parameter $A$ represents an intrinsic property of the FRB source and/or the physical process responsible for the emission of radiation. We conducted our analysis for $0.07\leq A\leq 0.2$, and since these values demonstrated similar trends, we selected $A=0.1$ for all of our subsequent plots. 

We adopt a frequency-dependent intrinsic duration following previous studies and the TRDM framework, in which the sub-burst duration varies inversely with observing frequency as $t_\mathrm{w}=t_\mathrm{w,0}/\nu$, with $t_\mathrm{w,0} \approx 1.5\,(\mathrm{ms \cdot GHz})$ \citep{Brown2024}. Frequency and duration are therefore intrinsically coupled: as we vary the central frequency over the range $0.4$ GHz to $8$ GHz, the duration varies according to this relation. In the case of ultra-FRBs, discussed in Section~\ref{sec:ultra_nar}, we reduce the scaling constant $t_\mathrm{w,0}$ to reflect their shorter intrinsic timescales. This relation also describes the mapping between duration and frequency in our complementary plots: shorter-duration bursts in the sub-burst slope–duration plot correspond to higher-frequency sub-bursts in the sub-burst slope–frequency plot, while longer-duration sub-bursts map to lower frequencies. Following the findings of \citet{Houde2019}, \citet{Chamma2023}, and \citet{Brown2024}, the bandwidth of the sub-burst is set to $\Delta \nu=0.14\,\nu_0$~(GHz), where $\nu_0$ is the central frequency of the sub-burst. Additionally, we have chosen a scattering index of $n=4.0$, although simulations with $n=4.4$ for the Kolmogorov spectrum also yield similar outcomes.

We also emphasize that we plot the negative of the sub-burst slope in all our figures. This approach allows for effective visualization of the sub-burst slope across a wide range of frequencies and durations using logarithmic scales (i.e., the sub-burst slope is intrinsically negative). However, as discussed in the forthcoming sections, there are instances where the sub-burst slope becomes positive due to excessive residual dispersion. In such cases, positive values are omitted from the plots due to the logarithmic scaling of the axes. 

\begin{figure}
 \includegraphics[width=\columnwidth ]{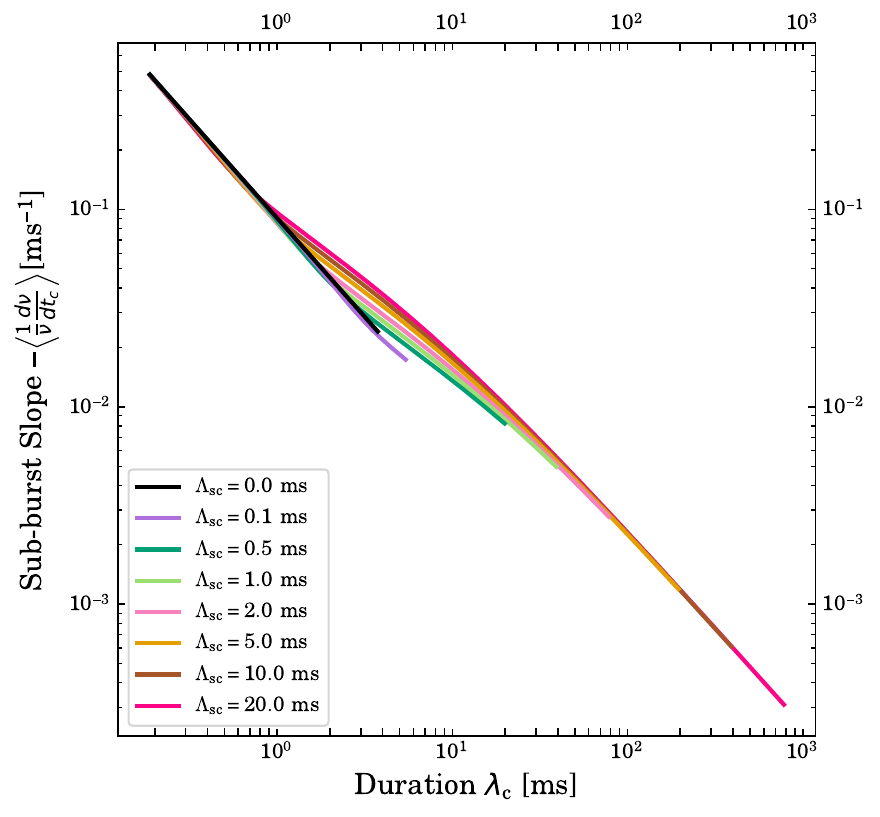}
 \caption{The relationship between the (negative of the) frequency-normalized sub-burst slope (Equation \ref{eq:slp_law_scat}) and the duration (Equation \ref{eq:lam_scat}) at the center frequency for different values of scattering timescales ($\Lambda_\mathrm{sc}$). The black line shows the ideal law without scattering, given by Equation (\ref{eq:slp_law_tc}).} 
 \label{fig:slp_law_1}
\end{figure}

\subsection{Effects of Scattering on the Sub-burst Slope Law}
\label{sec:scat_res}
To evaluate the sub-burst slope under varying scattering conditions, we perform computations across eight different scattering timescales, including the scenario of no scattering ($\tau_{\mathrm{sc}} = 0$ ms). As our focus here is exclusively on scattering, no other frequency-dependent propagation effects are present. 

Figure \ref{fig:slp_law_1} presents the (negative of the) center frequency normalized sub-burst slope law for various scattering timescales, determined using Equation (\ref{eq:slp_law_scat}). The solid black line represents the ideal sub-burst slope law without scattering and serves as the baseline for comparison. Our computation spans an intrinsic duration of $0.19 \ \mathrm{ms}\lesssim t_\mathrm{w}\leq 3.75\ \mathrm{ms}$ for the chosen parameters. For durations below $\sim 1 \ \mathrm{ms}$, the curves for different $\Lambda_\mathrm{sc}$ values are indistinguishable from the solid black curve, indicating that at shorter durations, the curves behave as though they are unscattered. As the sub-burst duration increases, curves corresponding to larger $\Lambda_\mathrm{sc}$ diverge first from the baseline. This is followed by a mild non-linear transition, after which the trajectories run parallel to the unscattered law but with an upward offset, indicative of a larger proportionality factor. Curves with minimal scattering ($\Lambda_\mathrm{sc} = 0.1,\ 0.5$ ms) show only slight deviations from the ideal case. In essence, the plots indicate that multipath scattering preserves the underlying inverse dependence, with a slight vertical offset between the short- and long-duration regimes.

\begin{figure}
 \includegraphics[width=\columnwidth ]{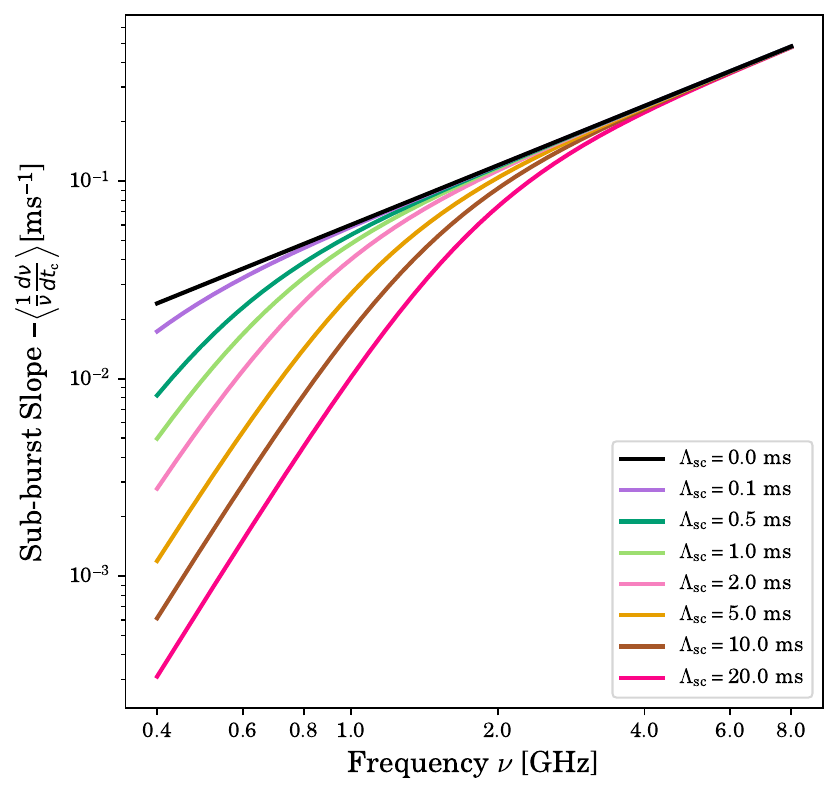}
 \caption{The (negative of the) frequency-normalized sub-burst slope against the sub-burst central frequency in range of 0.4 GHz to 8 GHz for different scattering timescales ($\Lambda_\mathrm{sc}$). The solid black line corresponds to the unscattered law (Equation \ref{eq:sub_slope_tc_freq}). 
}
 \label{fig:slp_law_freq}
\end{figure}

Our relation for the centroid-based sub-burst slope can be expressed explicitly as a function of the central frequency by substituting the expressions of the different timescales in the integrand of Equation (\ref{eq:slp_law_tc}),
\begin{align}
     \frac{1}{\nu}\frac{d\nu}{dt_{\mathrm c}}=-\frac{1}{t_\mathrm{w}(1 + 1/A)} = -C_1 \nu,
    \label{eq:sub_slope_tc_freq}
\end{align}
where $C_1 = [t_\mathrm{w,0} (1+1/A)]^{-1}$ is a constant. This is plotted using a solid black line in Figure \ref{fig:slp_law_freq}, where we show the frequency behavior of the sub-burst slope law. As evident from the plot, all curves align closely with the ideal unscattered case (solid black line) at higher frequencies ($\nu > 4.0\, \mathrm{GHz}$). In contrast, the curves exhibit clear deviations from the ideal law at lower frequencies ($\nu < 4.0\, \mathrm{GHz}$). The sub-burst slope drops significantly (for $\Lambda_{\mathrm{sc}}\geq 2 \, \mathrm{ms}$ at frequencies below $\sim 2$~GHz) and proportionately to the amount of scattering present in the sub-bursts. The curves with negligible amounts of scattering ( e.g., $\Lambda_{\mathrm{sc}}= 0.1\, \mathrm{ms}$) deviate only slightly from the unperturbed law down to the lowest frequencies.

\subsection{Effects of Residual Dispersion on the Sub-burst Slope Law}
\label{sec:DM_1}

\begin{figure}
 \includegraphics[width=\columnwidth ]{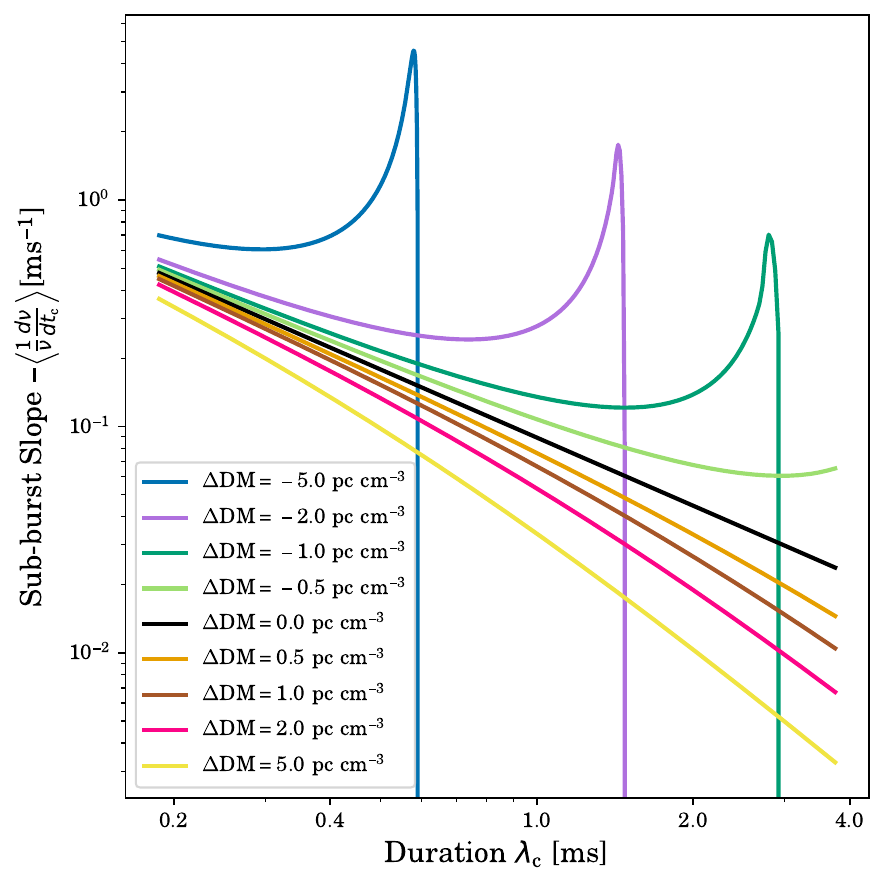}
 \caption{Sub-burst slope vs.~sub-burst duration for residual dispersion measures in the interval  $-5.0$~\pcm $\leq\Delta \mathrm{DM} \leq$ to  $5.0$~\pcm. The solid black curve represents the undispersed reference given by Equation (\ref{eq:slp_law_tc}). For all curves, $\tau_\mathrm{sc}=0$, and therefore, $\lambda_\mathrm{c}=t_\mathrm{w}(\nu_0)$. 
 } 
 \label{fig:slp_law_dm}
\end{figure}

Figure \ref{fig:slp_law_dm} presents the frequency-normalized sub-burst slope evaluated from Equation (\ref{eq:slp_law_dm}) for residual dispersion measures ($\Delta \mathrm{DM}$) ranging from $-5.0$~\pcm~to $+5.0$~\pcm. Multipath scattering is absent ($\tau_\mathrm{sc}=0.0 \ \mathrm{ms}$). Two systematic trends become evident. First, over-dedispersion ($\Delta \mathrm{DM}< 0$) aggressively steepens the sub-burst slope, and at sufficiently negative $\Delta \mathrm{DM}$ values, the drift reverses sign. This portion is omitted due to the logarithmic scaling of the axes. Second, under-dedispersion ($\Delta \mathrm{DM}> 0$) decreases the sub-burst drift $\left | d\nu/dt_\mathrm{c}\right |$ and shifts the curve below the baseline. In both cases, the vertical offsets are governed by the magnitude of the residual dispersion ($\left | \Delta \mathrm{DM} \right |$) present in the sub-burst.

\begin{figure}
 \includegraphics[width=\columnwidth ]{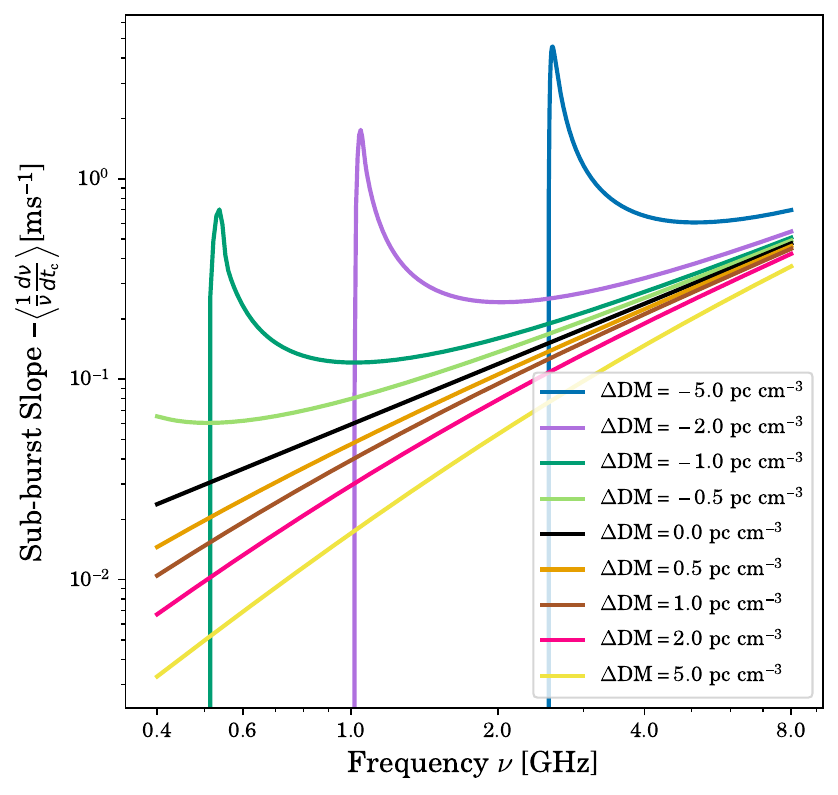}
 \caption{The frequency-normalized sub-burst slope vs.~frequency for $\Delta\mathrm{DM}$ in the range of $-5.0$~\pcm \ to  $5.0$~\pcm. The solid black curve represents the baseline law for sub-bursts with no residual dispersion. 
 } 
 \label{fig:slp_law_dm_freq}
\end{figure}

We also examine the sub-burst slope as a function of frequency in Figure \ref{fig:slp_law_dm_freq}. The inverse square dependence of DM on frequency, which disproportionately affects lower frequencies, is clearly apparent from this figure. We observe trends similar to those in Figure \ref{fig:slp_law_dm} with over-dedispersed bursts lying above the (solid black) baseline. When the applied over-dedispersion is large, the sub-burst slope changes sign and crosses through zero at relatively higher frequencies. For instance, when $\Delta \mathrm{DM} = -5.0 $~\pcm, the sub-burst slope changes sign at frequency $\nu \gtrsim 2.0 \ \mathrm{GHz}$. As the applied over-dedispersion decreases, for example $\Delta \mathrm{DM} \gtrsim -2.0$~\pcm, the zero-crossing migrates toward lower frequencies, $\nu <  2.0 \ \mathrm{GHz}$. Conversely, a positive residual DM (under-dedispersion), shifts the curves below the baseline by a magnitude that grows with $\Delta \mathrm{DM}$. This downward offset becomes progressively more pronounced towards the low-frequency end of the band.

\subsection{Combined Effects of Scattering and Dispersion on the Sub-burst Slope Law}
\label{sec:scat_dm}
Here, we plot the sub-burst slope law, as described in Equation (\ref{eq:slp_law_joint}). We adopt a single scattering timescale of $\Lambda_{\mathrm{sc}} = 2.0$~ms and compute results for $-5.0$~\pcm$ \leq \Delta\mathrm{DM} \leq +5.0$~\pcm.

\begin{figure}
 \includegraphics[width=\columnwidth ]{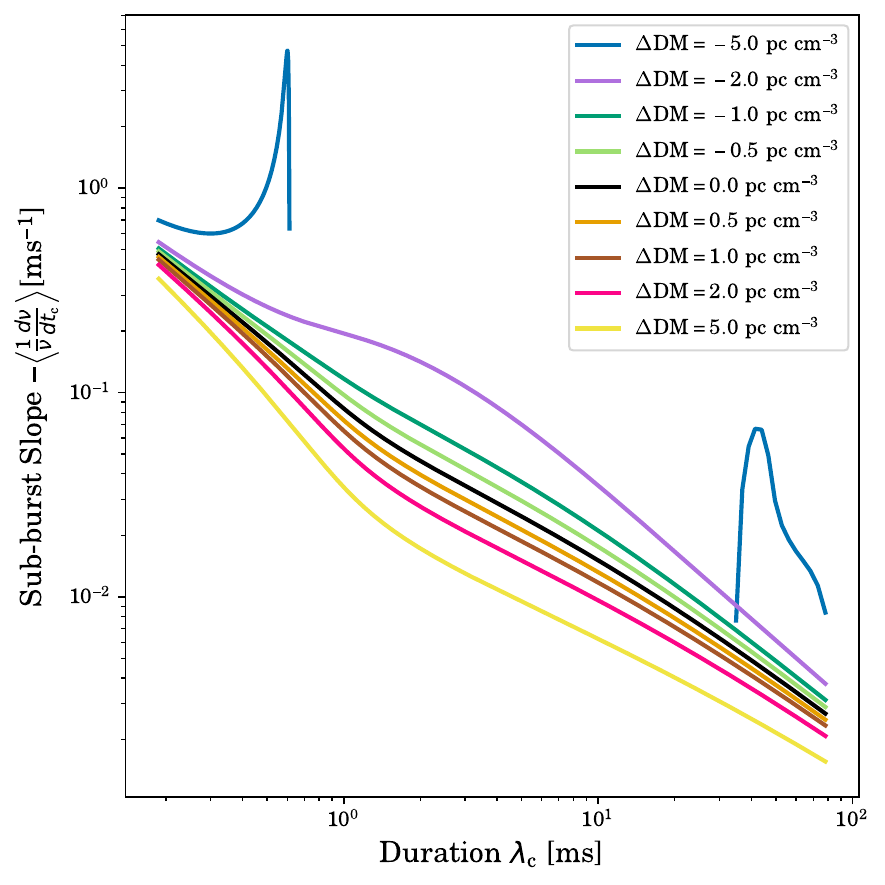}
 \caption{The frequency-normalized sub-burst slope against the duration for $-5.0\;\mathrm{pc\,cm^{-3}}\leq\Delta\mathrm{DM}\leq +5.0\;\mathrm{pc\,cm^{-3}}$. All curves have a fixed scattering timescale of $\Lambda_\mathrm{sc}=2$~ms. The solid black curve represents the law with no residual dispersion.}
 \label{fig:slp_law_dm_scat}
\end{figure}

Figure \ref{fig:slp_law_dm_scat} depicts the behavior of the sub-burst slope as a function of duration across the chosen $\Delta \mathrm{DM}$ range. We observe complex curve profiles, particularly for over-dedispersed sub-bursts. In the figure, the solid black line corresponds to the scattering-exclusive model with $\Delta \mathrm{DM} = 0.0$~\pcm, and thus serves as a baseline for comparison. Except for the extreme over-dedispersed case ($\Delta \mathrm{DM} = -5.0$~\pcm), every other curve retains the same morphology as the baseline curve: two limiting regions that scale as $\lambda_\mathrm{c}^{-1}$ at shorter and longer durations coupled by a broader nonlinear regime that is now severely modulated by dispersion. Over-dedispersion increases the magnitude of sub-burst slope, causing them to commence and deviate upwards from the baseline curve. Under-dedispersion reduces the slope at the outset, offsetting the curves to lie below the baseline. For $\Delta \mathrm{DM} = -5.0$~\pcm, the slope crosses zero in the transition region, and this segment is therefore absent in our plots. At larger durations, we see this curve re-emerging and converging onto the common asymptote followed by the other curves in this regime.

\begin{figure}
 \includegraphics[width=\columnwidth ]{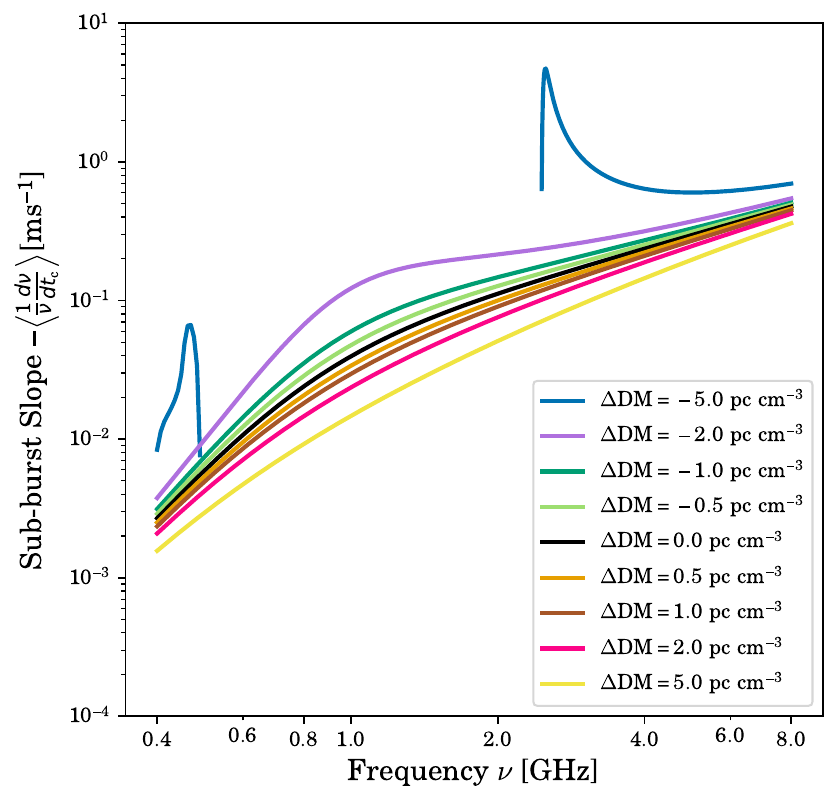}
 \caption{Same as Figure \ref{fig:slp_law_dm_scat} but as a function of the center frequency in the range $0.4 \ \mathrm{GHz} \leq \nu \leq 8.0 \ \mathrm{GHz} $. 
 }
 \label{fig:slp_law_freq_dm_scat}
\end{figure}

We observe complementary behavior in Figure \ref{fig:slp_law_freq_dm_scat}, which presents the same sub-burst slope as a function of frequency. At high frequencies, curves lie closer to the baseline $\Delta \mathrm{DM}=0.0$~\pcm, with the exception of the $\Delta \mathrm{DM}=-5.0$~\pcm~case. As the frequency decreases, the curves enter a transition region, displaying upward or downward offsets depending on the value of $\Delta \mathrm{DM}$. At the lowest frequencies, the curves converge toward the baseline. This resulting convergence at low $\nu$ is a distinctive feature of this joint model, absent from the purely scattered (Figure \ref{fig:slp_law_freq}) and purely dispersed (Figure \ref{fig:slp_law_dm_freq}) cases.

\subsection{Ultra-FRBs and Propagation Effects}
\label{sec:ultra_nar}
Multiple studies, including \citet{Nimmo2022}, \citet{Hewitt_2023}, and \citet{Snelders2023}, have reported the detection of ultra-FRBs: sub-bursts from repeating sources with durations ranging from nanoseconds to microseconds. We anticipate that such ultra-narrow sub-bursts are particularly susceptible to propagation effects, which may remain significant even at higher frequencies and shorter scattering timescales. To test this, we adjust our parameters to model sub-bursts with durations of $t_\mathrm{w,0}=50 \ \mathrm{\mu s} $ at 1~GHz (and scaling inversely with frequency; see Equation \ref{eq:tw}). The same analysis presented in Sections \ref{sec:scat_ex} and \ref{sec:dm_ex} is then implemented to study the resulting sub-burst slope law. 

\begin{figure}
 \includegraphics[width=0.8\columnwidth ]{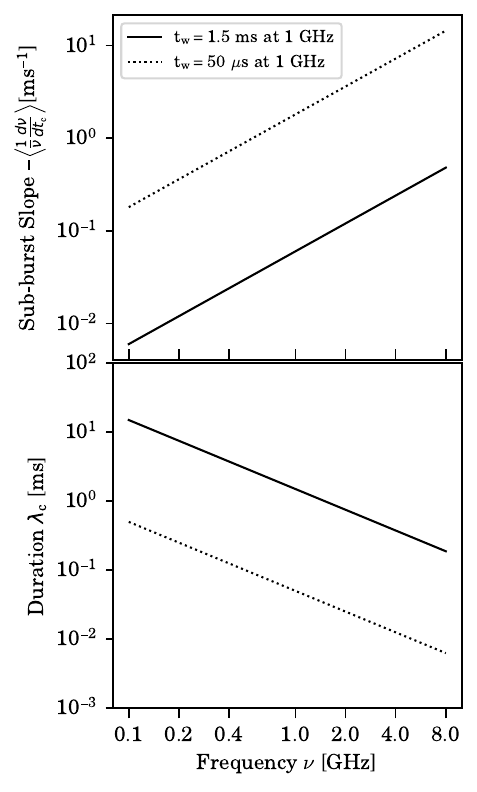}
 \caption{The (negative of the) frequency-normalized sub-burst slope (top panel) and duration (bottom panel) are plotted as a function of frequency for standard FRBs with a duration of 1.5~ms at 1~GHz (solid line) and for ultra-FRBs with a duration of $50\,\mathrm{\mu s}$ at 1~GHz (dotted line). No propagation effects are present. 
 }
 \label{fig:comb_plt}
\end{figure}

In Figure \ref{fig:comb_plt}, we present the (negative of the) normalized sub-burst slope (top plot) as well as the duration (bottom plot) as functions of frequency for the ideal case without any propagation effects, comparing ultra-fast FRBs (dotted line) with standard FRBs (solid line). The significantly shorter durations of ultra-FRBs  position them distinctly on duration-frequency plots, effectively forming a separate family of sub-bursts. Their durations are a factor of 30 (i.e., $1.5\,\mathrm{ms}/50\,\mu\mathrm{s}$) lower than those of standard FRBs at all frequencies. At the same frequency, the sub-burst slope of ultra-fast FRBs is steeper. Despite these differences, both families follow the same sub-burst slope law, underscoring the robustness and universality of this relationship across different FRB populations.

\begin{figure}
 \includegraphics[width=\columnwidth ]{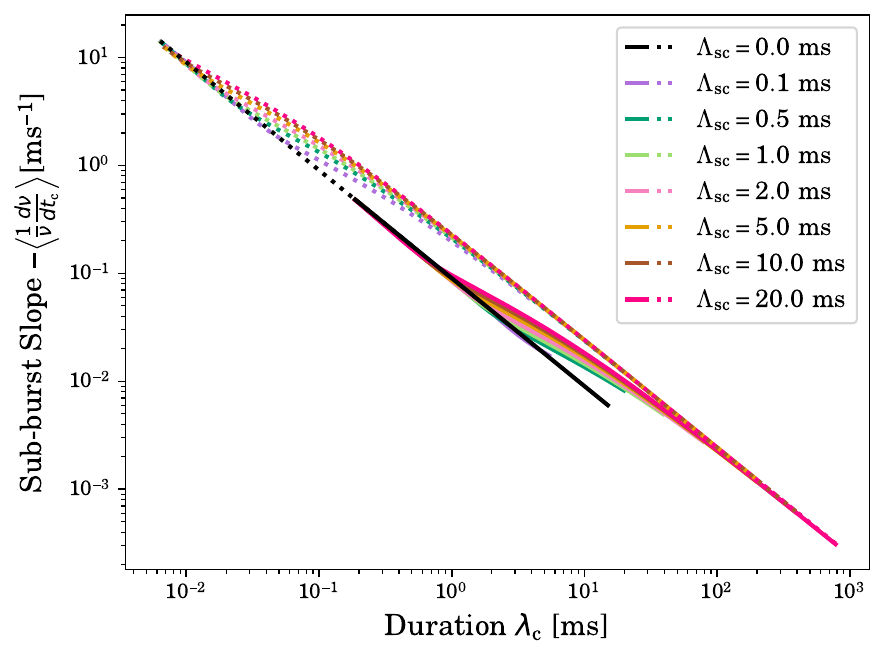}
 \caption{The (negative of the) frequency-normalized sub-burst slope as a function of duration with varying degrees of scattering for microsecond- and millisecond-duration sub-bursts. The solid black line represents the ideal sub-burst slope law in the absence of scattering. Solid colored lines correspond to millisecond-duration sub-bursts (i.e., 1.5~ms at 1~GHz), while dotted lines of the same colors represent microsecond-duration sub-bursts (i.e., $50\,\mathrm{\mu s}$ at 1~GHz). 
 }
 \label{fig:uf_slp_law_scat}
\end{figure}

For the scattering-exclusive case, our findings for ultra-FRBs, represented by dotted lines in Figure \ref{fig:uf_slp_law_scat}, are juxtaposed against our previous analysis of 1.5~ms duration sub-bursts at 1 GHz (solid lines) to facilitate a comparative visualization. While the overall behavior (as discussed in Section \ref{sec:scat_res}) is similar for both, the key distinction lies in their sensitivity to scattering. For ultra-FRBs, even modest scattering timescales ($\Lambda_\mathrm{sc} \sim 0.1 \ \mathrm{ms}$) significantly broaden the durations and attenuate the magnitude of the sub-burst slope. At sufficiently large durations (corresponding to lower frequencies), the scattered ultra-FRB curves converge toward, and become indistinguishable from, those of standard-width FRBs.

\begin{figure}
 \includegraphics[width=\columnwidth ]{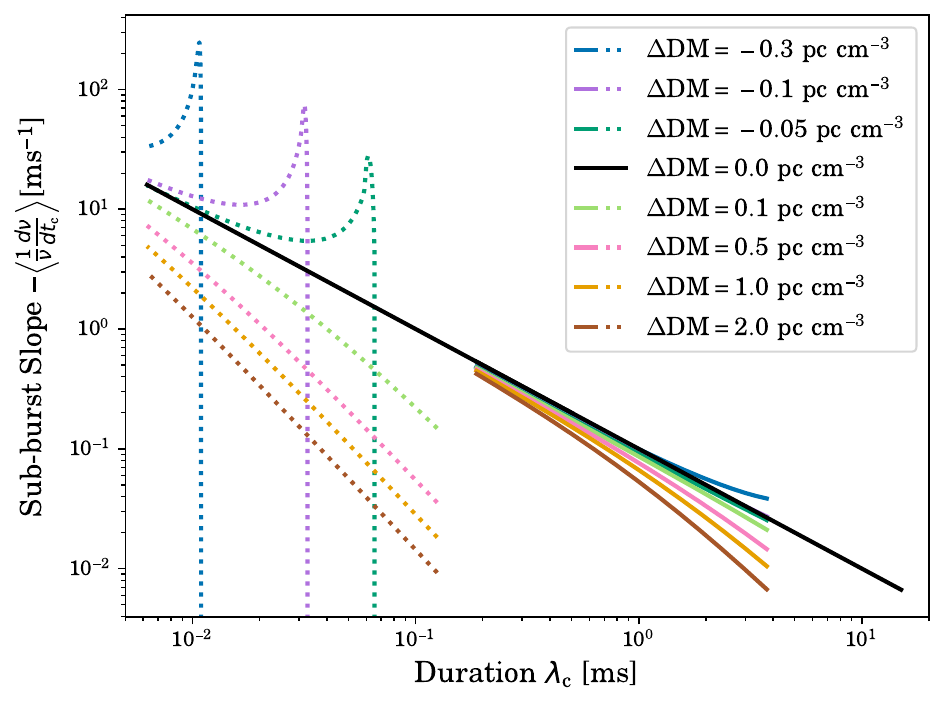}
 \caption{The sub-burst slope law as a function of duration for microsecond- and millisecond-duration sub-bursts for $-0.2$~\pcm $\leq\Delta\mathrm{DM}\leq$~$2.0$~\pcm. The solid black line represents the ideal sub-burst slope law in the absence of any residual dispersion. Solid colored lines correspond to millisecond-duration sub-bursts (i.e., 1.5~ms at 1~GHz), while dotted lines (following the same color scheme) represent microsecond-duration sub-bursts (i.e., $50\,\mathrm{\mu s}$ at 1~GHz).} 
 \label{fig:uf_slp_law_dm}
\end{figure}

In Figure \ref{fig:uf_slp_law_dm}, we present the sub-burst slope as a function of duration for ultra-FRBs over a restricted residual DM range of $-0.3$~\pcm \ to $2.0$~\pcm. While the overall behavior resembles the trends discussed in Section \ref{sec:DM_1}, we observe that smaller $\left | \Delta \mathrm{DM} \right |$ values result in more pronounced deviations in both the over- and under-dedispersed curves. In contrast, the standard FRBs stay relatively close to the baseline law ($\Delta \mathrm{DM} = 0.0$~\pcm) throughout this DM range.

\section{Discussion}
\label{sec:dis}
\subsection{Scattering Exclusive Analysis}
\label{sec:scat_dis}
The trends observed in Section \ref{sec:scat_res} can be understood from the interplay between the two contesting timescales, $t_\mathrm{w}$ and $\tau_\mathrm{sc}$, in the sub-burst slope law (Equation \ref{eq:slp_law_scat}). As each timescale follows a distinct frequency dependence, the behavior of the slope law can be divided into three regimes determined by their relative strengths:

\begin{enumerate}
    \item \textit{Scattering-subdominant Regime $t_\mathrm{w} \gg \tau_\mathrm{sc}$:} In the high-frequency regime, where the intrinsic width of the burst dominates over the scattering timescale, we can set $ t_\mathrm{c} \simeq t_\mathrm{D} + t_\mathrm{w}$. The sub-burst slope law reduces to Equation~(\ref{eq:slp_law_tc}), and its frequency dependence follows Equation~(\ref{eq:sub_slope_tc_freq}). The (unaveraged) sub-burst slope simplifies to 
    \begin{align}
    \frac{1}{\nu}\frac{d\nu}{dt_{\mathrm c}}
    \approx -
         \frac{A}{ t_{\mathrm w}},
    \label{eq:slp_law_weak}
    \end{align}
    since $A\ll 1$, and the duration reduces to $\lambda(\nu) \approx t_\mathrm{w}$. In this limit, the inverse sub-burst slope–duration scaling is preserved, the slope-frequency relation remains linear, and the curves coincide with the unscattered baseline, as seen in Figures \ref{fig:slp_law_1} and \ref{fig:slp_law_freq}. A fit to the slope law in this regime therefore recovers the intrinsic sub-burst slope parameter $A$. As $\nu$ decreases (or as $\lambda_\mathrm{c}$ increases), the curves with the largest $\Lambda_\mathrm{sc}$ are first to depart from the baseline, while those with smaller $\Lambda_\mathrm{sc}$ remain in this regime longer.

    \item {\textit{Scattering-dominated Regime $\tau_\mathrm{sc} \gg t_\mathrm{w}$:} As the frequency decreases, the scattering timescale starts to take precedence. In this limit, the duration is governed by the exponential tail, $\lambda \approx \tau_\mathrm{sc}$. Approximating the centroid to $ t_\mathrm{c} \simeq \tau_\mathrm{sc}$ and substituting its frequency scaling into Equation (\ref{eq:slp_law_scat}) yields the (unaveraged) scattering-dominated asymptote:
    \begin{align}
    \frac{1}{\nu}\frac{d\nu}{dt_{\mathrm c}} &\approx -\frac{1}{ n\tau_\mathrm{sc}}\nonumber\\ 
         &\approx -\frac{1}{ n\Lambda_\mathrm{sc}} \left( \frac{\nu}{1\ \mathrm{ GHz}} \right )^n. 
    \label{eq:slp_law_strong}
    \end{align}
    The inverse scaling with duration is thus preserved in the sub-burst slope–duration relationship. However, as seen in Figure \ref{fig:slp_law_1}, the scattered curves converge to a line parallel to the unscattered curve but displaced upward by an amount equal to $\left[n^{-1}-A/\left(1+A\right)\right]/\lambda_\mathrm{c}$. Thus, scattered bursts would yield larger estimates of $A$ compared to unscattered bursts. Similarly, in Figure \ref{fig:slp_law_freq}, the sub-burst slope–frequency relationship transitions to a new power-law scaling, characterized by an exponent equal to $n$, with its vertical offset determined by both $n$ and the magnitude of the scattering timescale ($\Lambda_\mathrm{sc}$).}
    
    \item  \textit{Intermediate Regime $t_\mathrm{w} \sim \tau_\mathrm{sc}$:} In the region where the two timescales are comparable, the sub-burst slope curves exhibit noticeable non-linearity. In Figure \ref{fig:slp_law_1}  this appears as a gentle curvature separating the scattering-subdominant and scattering-dominant asymptotes. No single power law approximation can accurately describe this regime, as both $t_\mathrm{w}$ and $\tau_\mathrm{sc}$ contribute comparably to the sub-burst slope and duration calculations. A similar trend appears in Figure \ref{fig:slp_law_freq} where the slope shows a softer decline in the mid-frequency region before approaching the scattering-dominated asymptote.
    
\end{enumerate}

To better illustrate the different regimes, we examine the case of $\Lambda_\mathrm{sc} = 2.0$ ms at 1 GHz used in our analysis. The unscattered slope across the frequency range is $\approx -0.067\ \mathrm{ms}^{-1}$. At 2 GHz, we obtain $t_\mathrm{w}= 0.75$~ms, $\tau_\mathrm{sc} = 0.125$~ms, $\lambda_\mathrm{c} \simeq 0.76$~ms. Since $t_\mathrm{w} \gg \tau_\mathrm{sc}$, this lies in the scattering-subdominant regime, and the normalized slope ($\approx -0.067~\mathrm{ms}^{-1}$) is consistent with the unscattered relation, corresponding to $A\simeq0.1$.  At 0.6 GHz, the parameters are $t_\mathrm{w}= 2.5$~ms, $\tau_\mathrm{sc} \simeq 15.43$~ms, $\lambda_\mathrm{c} \simeq 15.63$~ms. Here, $\tau_\mathrm{sc} \gg t_\mathrm{w}$, placing this sub-burst in the scattering-dominated regime. The normalized slope is $\simeq -0.0156\ \mathrm{ms}^{-1}$, close to the approximation from Equation~(\ref{eq:slp_law_strong}) ($\simeq -0.0162\ \mathrm{ms}^{-1}$), demonstrating that the slope remains inversely proportional to duration. The magnitude of the slope differs markedly from both the unscattered baseline and the scattering-subdominant regime. Furthermore, the value of $A$ in this regime increases to 0.25 ($A \propto n^{-1}$), significantly higher than in the subdominant case.

\subsection{DM-exclusive Analysis}
In the analysis presented in Section \ref{sec:dm_ex}, the sub-burst slope is inversely proportional to $t_\mathrm{D} + t_\mathrm{w} + 2\Delta t_\mathrm{DM}$. Increasingly negative residual dispersion values make the $2\Delta t_\mathrm{DM}$ term more negative (see Equation \ref{eq:dm_1}), resulting in higher slope magnitudes. Thus, in over-dedispersed curves, we observe an upward vertical offset across all  frequencies (and durations), with the magnitude of the offset increasing with decreasing observing frequency and increasingly negative $\Delta\mathrm{DM}$. Because both $t_\mathrm{D}$ and $t_\mathrm{w}$ scale as $\nu^{-1}$ and $\Delta t_\mathrm{DM}$ as $\nu^{-2}$, their relative contributions introduce a non-linear upward curvature. This is evident for all $\Delta \mathrm{DM} < 0.0 $~\pcm~curves in Figures \ref{fig:slp_law_dm} and \ref{fig:slp_law_dm_freq}. Once $2\left | \Delta t_\mathrm{DM} \right | > t_\mathrm{D} + t_\mathrm{w}$, the sub-burst slope changes sign. The positive values of the sub-burst slope are considered non-physical and are not visible on our log-log plot.

Under-dedispersed sub-bursts, which result from insufficient DM correction (i.e., when $\Delta\mathrm{DM}>0$), exhibit shallower slopes. In our sub-burst slope law (Equation \ref{eq:slp_law_dm}), it follows immediately that any positive residual dispersion increases the denominator $t_\mathrm{D} + t_\mathrm{w} + 2\Delta t_\mathrm{DM}$, thereby reducing the magnitude of the slope and producing a downward vertical offset. Furthermore, Figures \ref{fig:slp_law_dm} and \ref{fig:slp_law_dm_freq} confirm that for modest offsets, $-1\,\mathrm{pc \ cm^3}\le \Delta \mathrm{DM}\le1$~\pcm, the departure from the unperturbed law remains negligible at  $\nu > 1 \ \mathrm{GHz}$ (consistent with Figure 3 of \citealt{Chamma2023}). The true extent of any deviation, however, is contingent upon the frequency range of the sub-bursts considered and the magnitude of $\Delta \mathrm{DM}$.

\subsection{Joint Scattering-DM Analysis}
 Here, the sub-burst slope law (Equation~\ref{eq:slp_law_joint}) scales inversely with the composite timescale $t_\mathrm{D} +t_\mathrm{w}  + 2 \Delta t_\mathrm{DM} + n\tau_\mathrm{sc}$. At high frequencies, we have $t_\mathrm{D} +t_\mathrm{w} > 2 \Delta t_\mathrm{DM} + n\tau_\mathrm{sc}$, with residual dispersion dominating over scattering (i.e., $ 2 \Delta t_\mathrm{DM} > n\tau_\mathrm{sc}$) and shifting the curves vertically relative to the undispersed baseline (see Figures \ref{fig:slp_law_dm_scat} and \ref{fig:slp_law_freq_dm_scat}). At low frequencies, scattering grows rapidly and takes over once $n\tau_\mathrm{sc}  > t_\mathrm{D} +t_\mathrm{w} + 2 \Delta t_\mathrm{DM} $. Between these limits, the interplay of the timescales produces non-linear deviations whose form depends on the type and magnitude of the residual dispersion.

For severe over-dedispersion ($\Delta\mathrm{DM}=-5.0$~\pcm), the curves initially follow the residual dispersion-only solution (Figures ~\ref{fig:slp_law_dm} and \ref{fig:slp_law_dm_freq}) until the slope changes sign. As the frequency decreases, scattering overtakes and the track re-emerges, joining the scattering-dominated asymptote of Equation (\ref{eq:slp_law_strong}). Moderately over-dedispersed sub-bursts ($\Delta\mathrm{DM}=-2.0$~\pcm~and $-1.0$~\pcm) behave similarly: negative residual dispersion steepens the slope, but scattering suppresses the turnover and redirects the track toward its asymptote. For small residual dispersion values ($\Delta \mathrm{DM}=\pm0.5$~\pcm), the influence of $\Delta\mathrm{DM}$ remains subtle. The duration is set by $t_\mathrm{w}$ or $\tau_\mathrm{sc}$ at high and low frequencies, respectively. So these curves lie close to the solid black line. Under-dedispersion ($1.0$~\pcm$\leq \Delta \mathrm{DM}\leq$$5.0$~\pcm) monotonically increases the composite timescale, reducing the slope. These tracks converge to the baseline at the high- and low-frequency limits, with mild nonlinear deviations in between due to the contrasting frequency dependencies of the timescales.

\subsection{Ultra-FRBs}
 Figure \ref{fig:uf_slp_law_scat} demonstrates that a scattering timescale of only $0.1$ ms at 1 GHz is sufficient to rapidly drive an ultra-FRB ($t_\mathrm{w} \sim 50 \  \mu\mathrm{s}$ at $1 \ \mathrm{GHz}$) from the subdominant to the dominant scattering regime. This occurs because $t_\mathrm{w}$ (and therefore $t_\mathrm{D}$) is small for this family of bursts, so even a minimal contribution from scattering outweighs the other terms in the denominator of Equation~(\ref{eq:slp_law_scat}). When the same $\Lambda_\mathrm{sc} = 0.1$~ms is applied to a standard FRB ($t_\mathrm{w} \sim 1.5$~ms at 1~GHz), it produces only a modest deviation from the unscattered law (see Figure~\ref{fig:slp_law_1}). This implies that scattered ultra-FRBs retain the inverse dependence on duration but yield a higher scaling constant, as $A \propto n^{-1}$ in the scattering-dominated regime. At longer durations (lower frequencies), the sub-burst slope–duration relation is governed almost entirely by scattering for both ultra- and standard-width FRBs. The intrinsic width information is largely erased as these two regimes converge toward the same scattering asymptote. It is important to note that scattered ultra-FRBs with measured durations $\lambda_\mathrm{c} \gtrsim 300~\mu\mathrm{s}$ in Figure~\ref{fig:uf_slp_law_scat} become observationally indistinguishable from standard-width FRBs and would therefore be classified as standard FRBs. 

Ultra-FRBs are likewise hypersensitive to residual dispersion, as demonstrated in Figure \ref{fig:uf_slp_law_dm}. Even at $\Delta \mathrm{DM}=-0.05$~\pcm, the $2\Delta t_\mathrm{DM}$ term in the sub-burst slope law (Equation~\ref{eq:slp_law_dm}) dominates over $t_\mathrm{w} + t_\mathrm{D}$, producing a stronger non-linear upward curvature. In the case of under-dedispersion, $2\Delta t_\mathrm{DM}$ adds to $t_\mathrm{w} + t_\mathrm{D}$, reducing the sub-burst slope and shifting the curves downward, while the inverse duration scaling remains unchanged.

These results underscore the level of precision required when analyzing ultra-FRBs. Even minute residual dispersion or scattering effects that have a negligible impact on a millisecond-duration sub-burst can significantly alter the spectro-temporal characteristics of an ultra-FRB in view of its shorter duration. The DM estimates should be treated with caution unless either \textit{i)} the bursts are observed and dedispersed at sufficiently high frequencies, where scattering contributions become negligible, or \textit{ii)}  the accuracy of the dedispersion has been confidently verified to within $\left | \Delta \mathrm{DM} \right | < 0.05$~\pcm.

\section{Summary}
\label{sec:sum}

    We present a systematic analysis of propagation-induced modifications to the sub-burst slope law in both standard and ultra-FRBs. Scattering preserves the inverse slope–duration scaling in the asymptotic high- and low-frequency regimes, with a narrow intermediate region exhibiting non-linear behavior. We further examine the impact of residual dispersion, both independently and in combination with scattering. Over-dedispersion steepens the sub-burst slope and can even reverse its sign, whereas under-dedispersion progressively flattens it. When combined with scattering, the effects of over-dedispersion are attenuated, whereas those of under-dedispersion are enhanced. Ultra-FRBs follow qualitatively similar trends but exhibit heightened sensitivity due to their intrinsically short durations.

    The implications of this analysis are twofold. First, high-frequency observations are essential for accurately recovering intrinsic burst properties, as they are less affected by propagation effects and provide a reliable reference point. Second, induced deviations from the expected sub-burst slope law, whether arising from scattering, residual dispersion, or both, can lead to misinterpretations, particularly for ultra-FRBs. Our analysis helps characterize the impact of propagation effects on the sub-burst slope law, refining its application in modeling and constraining the physical processes that govern both the emission mechanism and the intervening medium. These distortions can, in principle, be identified and disentangled through our framework (or its generalization to other sub-burst or scattering profiles), thereby aiding the recovery of the underlying physical relationships.

\section*{Acknowledgments}
M.H.’s research is funded through the Natural Sciences and Engineering Research Council of Canada (NSERC) Discovery Grant RGPIN-2024-05242. F.R.'s research is supported by the NSERC Discovery Grant RGPIN-2024-06346. We thank the anonymous referee whose constructive remarks prompted us to revisit our analysis, resulting in a refined and simplified framework.

\bibliography{refs}{}
\bibliographystyle{aasjournal}

\end{document}